\documentclass[onecolumn,preprintnumbers,amsmath,amssymb,aps,prd,nofootinbib,eqsecnum,showpacs]{revtex4}

\usepackage{graphicx,epsf, epsfig, amssymb}
\usepackage{bm}

\usepackage{amsmath}

\def\be{\begin{equation}}
\def\ee{\end{equation}}
\def\beq{\begin{eqnarray}}
\def\eeq{\end{eqnarray}}

\def\IL{\relax{\rm I\kern-.18em L}}

\begin{document}

\title{Classical instability of Kerr-AdS black holes and the issue of final state}

\author{Vitor Cardoso}
\email{vcardoso@phy.olemiss.edu} \affiliation{Department of Physics and Astronomy, The University of
Mississippi, University, MS 38677-1848, USA \footnote{Also at Centro de F\'{\i}sica Computacional, Universidade
de Coimbra, P-3004-516 Coimbra, Portugal}}

\author{\'Oscar J. C. Dias}
\email{odias@perimeterinstitute.ca} \affiliation{Perimeter
 Institute for Theoretical Physics,
 Waterloo, Ontario N2L 2Y5, Canada  \\
         and \\
 Department of Physics, University of Waterloo, Waterloo, Ontario
 N2L 3G1, Canada}

\author{Shijun Yoshida}
\email{shijun@waseda.jp} \affiliation{Science and Engineering, Waseda University, Okubo, Shinjuku, Tokyo
169-8555, Japan}

\date{\today}

\begin{abstract}
It is now established that small Kerr$-$Anti-de Sitter (Kerr-AdS) black holes are unstable against {\it scalar}
perturbations, via superradiant amplification mechanism. We show that small Kerr-AdS black holes are also
unstable against {\it gravitational} perturbations and we compute the features of this instability. We also
describe with great detail the evolution of this instability. In particular, we identify its endpoint state. It
corresponds to a Kerr-AdS black hole whose boundary is an Einstein universe rotating with the light velocity.
This black hole is expected to be slightly oblate and to co-exist in equilibrium with a certain amount of
outside radiation.
\end{abstract}


\maketitle

\section{Introduction}
Black holes in an anti-de Sitter (AdS) background have attracted a
great deal of attention over the last decade. One of the reasons
for this intense study is the AdS/CFT correspondence
\cite{maldacena,witten0}, stating that there is a duality between
supergravity on ${\rm AdS} \times {\cal M}$ (with ${\cal M}$ a
compact manifold) and an appropriate conformal field theory (CFT)
defined on the boundary of the AdS bulk. For instance, M-theory on
${\rm AdS}_4\times S^7$ is dual to a non-abelian superconformal
field theory in three dimensions, and type IIB superstring theory
on ${\rm AdS}_5\times S^5$ seems to be equivalent to a super
Yang-Mills (SYM) theory in four dimensions
\cite{maldacena,witten0} (for a review see
\cite{maldacenareview}). Of course the duality is not clearly and
unambiguously understood, so much of the work on the subject
involved building a catalog, or a dictionary taking us from one
description to the other. Some of the most important issues to
understand are the thermodynamic and classical stabilities of
these AdS black holes.

Research on the thermodynamic stability of AdS black holes started
two decades ago. It was found that Schwarzschild-AdS black holes
are subjected to the Hawking-Page phase transition
\cite{hawkingpage}: at low temperatures a thermal gas is a
globally stable configuration in the AdS background (because in
this background, as oppose to the flat one, a black hole can exist
only for a temperature above a critical value) but when one
increases the temperature above a critical value, the thermal bath
becomes unstable and collapses to form a black hole. The
interpretation of this transition in the dual CFT side was
provided in \cite{witten0}. The thermal AdS space at low
temperature corresponds, in the dual picture, to a confined
thermal SYM theory. When we heat up the thermal SYM it deconfines,
i.e., quarks can be seen in isolation. It is then natural to argue
that the Hawking-Page phase transition in the gravity side of the
duality corresponds to a confinement/deconfinement transition in
the CFT side \cite{witten0,aharonyetall}. A number of other
asymptotically AdS black holes have been studied, and their
thermodynamical properties thoroughly investigated: charged
\cite{louwin,pele,mi1,mi2,cejm1,cejm2,cck,hr}, rotating
\cite{cck,hr,berpar,hhtr}, and non-spherical
\cite{birm,caldklem,emp} black holes, and their interpretation in
the AdS/CFT duality context.

On the other hand, the classical stability of Kerr-AdS black holes
is not as completely understood. One might worry that any rotating
black hole in AdS space might be unstable against superradiance: a
rotating black hole amplifies some incoming waves, and since AdS
is a box-like spacetime, the amplified radiation would be
reflected back, and so on, leading to a black hole bomb
\cite{teubhb,blackbomb,AdSbomb}, or exponential growth of any
perturbation. Two major results were reported:

(i) large ($r_+>\ell$) Kerr-AdS are stable
\cite{hawkingreall,blackbomb,AdSbomb} (where $r_+$ is the horizon
radius  and $\ell$ is the cosmological length). From a
superradiance point of view, this is because the characteristic
modes of large AdS holes are very large and do not satisfy the
superradiant condition $\omega<m \Omega_{\rm H}$ (see
\cite{blackbomb,AdSbomb,bs} for details). Here $\omega$ and $m$
are the frequency and angular momentum of the mode, and
$\Omega_{\rm H}$ is the horizon velocity of the black hole.

(ii) small ($r_+<\ell$) Kerr-AdS black holes are unstable against
{\it scalar} perturbations \cite{AdSbomb}, precisely by a black
hole bomb mechanism. The value $r_+ = r_+^{\rm c}=\sqrt{a\ell}$,
which obviously depends on the rotation parameter $a$, plays a
critical role. It was shown \cite{AdSbomb} that when, and only
when, $r_+ < r_+^{\rm c}$,  Kerr-AdS black holes exhibit a
classical (superradiant) instability. One of the purposes of this
paper is to show that small Kerr-AdS black holes are also unstable
against {\it gravitational} perturbations and therefore the metric
is itself unstable \footnote{Recently, while this work was on its
last stages, it was shown in \cite{KunduriLuciettiReall} that for
$D\geq 5$, small rotating AdS black holes are unstable against a
certain sector of gravitational perturbations.}.
 Now, the boundary of the Kerr-AdS black hole, where the dual CFT lives, is conformal to a rotating Einstein
universe. Remarkably, a Kerr-AdS black hole whose parameters satisfy the condition $r_+ = r_+^{\rm
c}=\sqrt{a\ell}$, is a black hole for which the corresponding Einstein universe on the boundary rotates
precisely with a critical value equal to the velocity of light. Black holes with $r_+ < r_+^{\rm c}$ correspond
to an Einstein universe that rotates faster than the speed of light, while for $r_+ > r_+^{\rm c}$ the Einstein
universe on the boundary rotates with velocity smaller than the light velocity. We will be able to follow the
black hole evolution induced by the classical instability, and to identify the precise final configuration that
describes the equilibrium endpoint of the instability. In particular, we will show that a small Kerr-AdS black
hole that starts with a given rotation and whose horizon radius is such that $r_+ < r_+^{\rm c}$, will lose
angular momentum and increase its radius until its radius and rotation satisfy the relation $r_+=r_+^{\rm
c}=\sqrt{a\ell}$. At this point the system becomes classically stable. The endpoint configuration is therefore
such that the Einstein universe that describes its boundary is rotating precisely with the speed of light.

In Section \ref{sec:instability}, we show that small four dimensional Kerr-AdS black holes are unstable against
gravitational perturbations, and we compute the instability timescale. From several arguments, presented in
\cite{blackbomb,AdSbomb} we expect higher dimensional Kerr-AdS black holes to be unstable as well. There is no
known formalism to handle gravitational perturbations of this class of geometries, but in
\cite{KunduriLuciettiReall} the authors focused on a particular class of gravitational perturbations (which
exist only for $D\geq5$) and showed that they too lead to instabilities. Using the results in
\cite{KunduriLuciettiReall} we compute in Appendix \ref{app:instddim} the instability timescale for higher
dimensional geometries. We close with Section \ref{sec:endpoint} where we discuss the possible final state of
this instability.

\section{\label{sec:instability} Classical instability of Kerr-A\lowercase{d}S black holes and its endpoint}

\subsection{Kerr-AdS black holes in four dimensions}
The metric of a four-dimensional Kerr-AdS black hole is
\begin{eqnarray} ds^2=-\frac{\Delta_r}{\rho^2}\left ( dt-\frac{a}{\Sigma}\sin^2\theta \,d\phi\right )^2
+\frac{\rho^2}{\Delta_r}\,dr^2
+\frac{\rho^2}{\Delta_{\theta}}\,d\theta^2
+\frac{\Delta_\theta}{\rho^2} \sin^2\theta\left (
a\,dt-\frac{r^2+a^2}{\Sigma} \,d\phi\right )^2 \,, \label{metric}
\end{eqnarray}
with
\begin{eqnarray}
& & \Delta_r=\left (r^2+a^2\right )\left (1+\frac{r^2}{\ell^2}
\right )-2Mr\,, \qquad  \Sigma=1-\frac{a^2}{\ell^2}
\nonumber \\
& &\Delta_{\theta}= 1-\frac{a^2}{\ell^2}\cos^2\theta\,, \qquad
\rho^2=r^2+a^2 \cos^2\theta  \,,
 \label{metric parameters}
\end{eqnarray}
and $\ell=\sqrt{-3/\Lambda}$ is the cosmological length associated with the cosmological constant $\Lambda$.
This metric describes the gravitational field of the Kerr-AdS black hole, with mass $M/\Sigma^2$, angular
momentum $J=a M/\Sigma^2$, and has an event horizon at $r=r_+$ (the largest root of $\Delta_r$). A
characteristic and important parameter of a Kerr black hole is the angular velocity of its event horizon given
by
\begin{equation}
\Omega_{\rm H}=\frac{a}{r_+^2+a^2}\left ( 1-\frac{a^2}{\ell^2}
\right ) \,.
  \label{Omega}
\end{equation}
The area of the black hole horizon is
\begin{eqnarray}
A=\frac{4\pi  (r_+^2+a^2)}{\Sigma} \,,
  \label{area}
\end{eqnarray}
and its temperature is
\begin{eqnarray}
T=\frac{r_+\left(1+a^2\ell^{-2}+3r_+^2\ell^{-2} -a^2 r_+^{-2}\right)}{4\pi (r_+^2+a^2)} \,.
  \label{temperature}
\end{eqnarray}
In order to avoid singularities, the black hole rotation is constrained to be
\begin{equation}
a<\ell \,.
  \label{upper rotation}
\end{equation}
\subsection{Wave equation for gravitational perturbations}
Gravitational perturbations of rotating black holes are best dealt with using Teukolsky's formalism
\cite{teukolsky}. The equation describing gravitational perturbations of Kerr-AdS black holes was obtained by
Chambers and Moss \cite{mosschamgia} and also Giammatteo and Moss \cite{mosschamgia}. The result is again a
system of two coupled ordinary differential equations, one radial and one angular.

To get the radial equation, start with Eq. (15) by Giammatteo and
Moss \cite{mosschamgia}, which can be written as
\begin{eqnarray} r^4\Delta_r\partial_r \left ( \frac{\Delta_r}{r^2}\partial_r R\right ) - r\Delta_r \left
(-\frac{r}{\Delta_r}(K_r+i\Delta_r')^2-\frac{2\Delta_r}{r}+\Delta_r'+
\lambda r-\frac{6r^3}{l^2} + 6ir^2\omega+irK_r'- r\Delta_r''
\right )R=0\,, \end{eqnarray}
where the primes stand for derivatives with respect to $r$, and
\begin{eqnarray}
K_r=-am\left( 1-\frac{a^2}{\ell^2}\right) +\omega (r^2+a^2) \,.
 \end{eqnarray}
 Defining $R=\frac{r}{\Delta_r}\Psi$ we get
\begin{eqnarray} \Delta_r \Psi''-\Delta_r'\Psi'-\left ( -\frac{K_r^2+2iK_r\Delta_r'}{\Delta_r} + \lambda-\frac{6r^2}{\ell^2}
+6ir\omega+iK_r' \right )\Psi=0 \,.\label{teukfinal}
\end{eqnarray}
 In the limit of asymptotically flat space ($\ell \rightarrow
\infty$), Eq. (\ref{teukfinal}) reduces to Teukolsky's
\cite{teukolsky} radial equation with spin $s=-2$. On the other
hand, for $a=0$ the following transformation
\begin{eqnarray} \Psi=\Delta_r\left (\frac{\Delta_r}{r^2}\partial_r+i\omega\right ) \left (\partial_r+\frac{i\omega
r^2}{\Delta_r}\right )r X(r)\,,\label{rwt}\end{eqnarray}
yields the Regge-Wheeler equation for $X(r)$, in Schwarzschild-AdS
spacetimes \cite{cardosogpert}. We note that the decomposition
used in \cite{mosschamgia} is of the form $e^{i\omega t+im\phi}$.
We adopted here the most common form $e^{i\omega t-im\phi}$ so
that waves co-rotating with the black hole have positive $m$. This
will be important in the sequel, to decide when and why there is
an instability.

The angular equation satisfies
\begin{eqnarray} \Delta_{\mu}^2 \partial_{\mu}^2 S+ \Delta_{\mu}(\partial_{\mu}\Delta_{\mu})\partial_{\mu}S
+\left [ -K_{\mu}^2-\Delta_{\mu}^2+\Delta_{\mu} \left
(\lambda-2\mu^2\Lambda+6\mu\omega
+2K_{\mu}+\partial_{\mu}\Delta_{\mu}\right ) \right
]S=0\,,\end{eqnarray}
where $S=S(\mu)$ and
\begin{eqnarray} \mu=a \cos \theta\,,\,\, \quad
K_{\mu}=-\left (1-\frac{a^2}{\ell^2}\right )am+(a^2-\mu^2)\,,\,\,\quad \Delta_{\mu}=(a^2-\mu^2)\left
(1-\frac{\mu^2}{\ell^2}\right ) \,,\end{eqnarray}
For small rotation and small cosmological constant, $\lambda \rightarrow (l-1)(l+2)$. The primes stand for
derivatives with respect to $\mu$. Likewise, this angular equation reduces to the usual spin-weight $2$
spherical harmonics when $a \rightarrow 0$ and it reduces to the spin-weight $2$ spheroidal harmonics
\cite{swsh} when $l \rightarrow \infty$.

\subsubsection{\label{sec:BH Near region}Solution in the near-region}
For small AdS black holes, $r_+/\ell \ll 1$, in the near-region, $r-r_+ \ll 1/\omega$, we can neglect the
effects of the cosmological constant, $\Lambda\sim 0$. Moreover, one has $r\sim r_+$, $r_+ \sim 2M$, and
$\omega a \sim 0$ (since $\omega\ll M^{-1}$ and $a\ll M$), and $\Delta_r \sim \Delta$ with
\begin{eqnarray}
\Delta=r^2+a^2-2Mr=(r-r_+)(r-r_-)\,.
 \label{near wave eqDelta}
\end{eqnarray}
The near-region radial wave equation can then be written as
\begin{eqnarray}
\Delta_r\partial_r^2 \Psi-\Delta_r'\partial_r \Psi-\left
((l-1)(l+2)-\frac{K_r^2+2iK_r\Delta_r'}{\Delta_r}\right )\,\Psi=0 \, .
 \label{near wave eq}
\end{eqnarray}
We find it convenient to express the last term in the following
form, near the event horizon:
\begin{eqnarray} K_r^2+2iK_r\Delta_r' \sim \left [ K_r+i(r_+-r_-)
\right ]^2+(r_+-r_-)^2\,,\end{eqnarray}
which using the explicit expression for $K_r$ can be shown to be, near $r_+$,
\begin{eqnarray} (K_r^2+2iK_r\Delta_r')^2 \sim (r_+-r_-)^2
\left [ (\varpi +i)^2+1\right ]\,.\end{eqnarray}
We have defined the superradiant factor
\begin{eqnarray}
\varpi \equiv(\omega-m\Omega_{\rm H})\frac{r_+^2+a^2}{r_+-r_-} \,,
 \label{superradiance factor}
\end{eqnarray}
which will play an important role in the instability. For
$\omega-m\Omega_{\rm H}<0$ this factor is negative, and we have
superradiance (see \cite{AdSbomb} for more details). It turns out
that it is also in this regime that the spacetime is unstable.

If one introduces a new radial coordinate,
\begin{eqnarray}
z=\frac{r-r_+}{r-r_-}\, , \qquad 0\leq z \leq 1\,,
 \label{new radial coordinate}
\end{eqnarray}
with the event horizon being at $z=0$, then the near-region radial wave equation can be written as
\begin{eqnarray}
& & \hspace{-0.5cm} z(1\!-\!z)\partial_z^2 \Psi-
(1\!+\!3z)\partial_z \Psi+ \left [(\varpi+i)^2+1\right ]
\frac{1\!-\!z}{z} \Psi - \frac{(l-1)(l+2)}{1\!-\!z} \Psi=0\,, \nonumber \\
& &
 \label{near wave eq with z}
\end{eqnarray}
Through the definition
\begin{eqnarray}
\Psi=z^{i \,\varpi} (1-z)^{l-1}\,F \,,
 \label{hypergeometric function}
\end{eqnarray}
the near-region radial wave equation takes the form
\begin{eqnarray}
& &  \hspace{-0.5cm} z(1\!-\!z)\partial_z^2 F+ {\biggl [} (-1+i\, 2\varpi)-\left [ 1+2l+ i\, 2\varpi \right
]\,z {\biggr ]}
\partial_z F -(l+1) \left [ l-1+ i \,2\varpi\right ] F=0\,.
 \label{near wave hypergeometric}
\end{eqnarray}
This wave equation is a standard hypergeometric equation \cite{abramowitz}, $z(1\!-\!z)\partial_z^2
F+[c-(a+b+1)z]\partial_z F-ab F=0$, with
\begin{eqnarray}
& & \hspace{-0.5cm} a=l-1+i\,2\varpi \,,  \qquad b=l+1 \,, \qquad c=-1+ i\,2\varpi \,.
 \label{hypergeometric parameters}
\end{eqnarray}
and its most general solution in the neighborhood of $z=0$ is $A\, z^{1-c} F(a-c+1,b-c+1,2-c,z)+B\,
F(a,b,c,z)$. Using (\ref{hypergeometric function}), one finds that the most general solution of the near-region
equation is
\begin{eqnarray}
 \hspace{-0.5cm} \Psi &=& A\, z^{2-i\,\varpi}(1-z)^{l-1}
F(a-c+1,b-c+1,2-c,z) +B\,z^{i\,\varpi}(1-z)^{l-1} F(a,b,c,z) \,.
 \label{hypergeometric solution}
\end{eqnarray}
The first term represents an ingoing wave at the horizon $z=0$, while the second term represents an outgoing
wave at the horizon. We are working at the classical level, so there can be no outgoing flux across the
horizon, and thus one sets $B=0$ in (\ref{hypergeometric solution}). This boundary condition also follows from
regularity requirements \cite{teukolsky} since $R\sim (r-r_+)^{-1}$ at the horizon for outgoing waves.

One is now interested in the large $r$, $z\rightarrow 1$, behavior of the ingoing near-region solution. To
achieve this aim one uses the $z \rightarrow 1-z$ transformation law for the hypergeometric function
\cite{abramowitz},
\begin{eqnarray}
& \hspace{-2cm} F(a\!-\!c\!+\!1,b\!-\!c\!+\!1,2\!-\!c,z)=
(1\!-\!z)^{c-a-b}
\frac{\Gamma(2-c)\Gamma(a+b-c)}{\Gamma(a-c+1)\Gamma(b-c+1)}
 \,F(1\!-\!a,1\!-\!b,c\!-\!a\!-\!b\!+\!1,1\!-\!z) & \nonumber \\
&  \hspace{-0.2cm}+ \frac{\Gamma(2-c)\Gamma(c-a-b)}{\Gamma(1-a)\Gamma(1-b)}
 \,F(a\!-\!c\!+\!1,b\!-\!c\!+\!1,-c\!+\!a\!+\!b\!+\!1,1\!-\!z)\,,
 \label{transformation law}
\end{eqnarray}
and the property $F(a,b,c,0)=1$. Finally, noting that when $r\rightarrow \infty$ one has $1-z = (r_+-r_-)/r$,
one obtains the large $r$ behavior of the ingoing wave solution in the near-region,
\begin{eqnarray}
R &\sim& A\,\Gamma(3-i\,2\varpi){\biggl [}
\frac{(r_+-r_-)^{-l-2}\,\Gamma(2l+1)}{\Gamma(l)\Gamma(l+2-i\,2\varpi)}\:
r^{l+2}+\frac{(r_+-r_-)^{l-1}\,\Gamma(-2l-1)}{\Gamma(-l)\Gamma(2-l-i\,2\varpi)}\: r^{1-l} {\biggr ]}.
 \label{near field-large r}
\end{eqnarray}

\subsubsection{\label{sec:BH Far region}Far-region wave equation
and solution}
In the far-region, $r-r_+ \gg M$, the effects induced by the black hole can be neglected ($a\sim 0$, $M \sim
0$, $\Delta_r \sim r^2[1+r^2/\ell^2]$) and the radial wave equation
 (\ref{teukfinal}) reduces to the wave equation of a
graviton in pure AdS background. We will first find the solution to the Regge-Wheeler equation and then use
(\ref{rwt}) to find the Teukolsky function itself. Regge-Wheeler's equation in pure AdS is \cite{cardosogpert}
\be (r^2+1)^2\partial_r^2 \Psi_{RW}+2r(r^2+1)\partial_r\Psi_{RW}+\left (\omega^2-\frac{l(l+1)}{r^2}\right
)\Psi_{RW}=0\,. \ee
An exact solution to this equation was found in \cite{cardosogpert} (see their Appendix):
\begin{eqnarray} \Psi_{RW}=\sqrt{x-1}\,x^{(l+1)/2} \left [ C F(\alpha,\beta,\gamma,x)+Dx^{1-\gamma}
F(1+\alpha-\gamma,1+\beta-\gamma,2-\gamma,x)
\right]\,,\end{eqnarray}
where
\begin{eqnarray} x=\frac{r^2}{r^2+1}\,,\,\,\quad \alpha=2+l-\omega\,,\,\,
\quad \beta=2+l+\omega\,,\,\,  \quad \gamma=3/2+l\,. \end{eqnarray}
Near infinity $x \sim 1$, we have, using known transformation laws for the hypergeometric functions,
\begin{eqnarray}  C F(\alpha,\beta,\gamma,x)
&\sim&C(1-x)^{-l-5/2}\frac{\Gamma[\gamma]\Gamma[\alpha+\beta-\gamma]}
{\Gamma[\alpha]\Gamma[\beta]}\,,\\
Dx^{1-\gamma} F(1+\alpha-\gamma,1+\beta-\gamma,2-\gamma,x)&\sim&
D(1-x)^{-l-5/2}\frac{\Gamma[2-\gamma]\Gamma[\alpha+\beta-\gamma]}
{\Gamma[\alpha-\gamma+1]\Gamma[\beta-\gamma+1]}\,. \end{eqnarray}
The boundary conditions appropriate for gravitational fields in AdS space have been discussed by a number of
authors \cite{moss}. We will adopt Dirichlet boundary conditions here, although we suspect the results hold for
more general boundary conditions (for instance, QNMs in AdS are weakly affected by boundary conditions
\cite{moss}; since there is a direct relation between the instability and QNMs in AdS \cite{blackbomb}, we
conjecture that the instability itself is weakly sensitive to these). These ``reflective'' boundary conditions
imply that
\begin{eqnarray} \frac{C}{D}=-\frac{\Gamma[2-\gamma]\Gamma[\alpha]\Gamma[\beta]}
{\Gamma[\gamma]\Gamma[\alpha-\gamma+1]\Gamma[\beta-\gamma+1]}=-\frac{\Gamma[1/2-l]\Gamma[2+l-\omega]\Gamma[2+l+\omega]}
{\Gamma[3/2+l]\Gamma[3/2-\omega]\Gamma[3/2+\omega]}\,.\end{eqnarray}

For small values of $r$, we have
\begin{eqnarray} \Psi_{RW}\sim iCr^{l+1}+iDr^{-l}\,. \end{eqnarray}
We can now use (\ref{rwt}), which for small values of $r$ takes
the form
\begin{eqnarray} \Psi \sim l(l+1)r\Psi_{RW}+2r^2\Psi_{RW}'\,. \end{eqnarray}
Thus,
\begin{eqnarray} \Psi \sim iC(l+1)(l+2)r^{l+2}+iDl(l-1)r^{-l+1} \label{far}\end{eqnarray}
Matching (\ref{far}) with  (\ref{near field-large r}) we get
\begin{eqnarray} \frac{(r_+-r_-)^{-2l-1}\Gamma[2l+1]\Gamma[-l]\Gamma[2-l-2i\varpi]}{\Gamma[l]\Gamma[l+2-2i\varpi]
\Gamma[-2l-1]}=-\frac{(l+1)(l+2)}{l(l-1)}\frac{\Gamma[1/2-l]\Gamma[2+l-\omega]\Gamma[2+l+\omega]}
{\Gamma[3/2+l]\Gamma[3/2-\omega]\Gamma[3/2+\omega]}
\,.\label{eigenfin}\end{eqnarray}
This is an eigenvalue equation for $\omega$. To show that instabilities are possible, we will expand this last
equation around the pure AdS value \cite{cardosogpert} which is
\begin{eqnarray} \omega_{\rm AdS}=\frac{ 2n+l+2}{\ell}\,, \label{adsfreq}\end{eqnarray}
with $n$ an integer. We thus set (see \cite{blackbomb,AdSbomb} for a more elaborate discussion on this)
\begin{eqnarray} \omega= \frac{2n+l+2}{\ell}+i\delta\,.\end{eqnarray}
First, we can use the following identities
\begin{eqnarray} \frac{\Gamma[-2l-1]}{\Gamma[-l]}=(-1)^{l+1}\frac{l!}{(2l+1)!}\,, \end{eqnarray}
\begin{eqnarray} \frac{\Gamma[l+2-2i\varpi]}{\Gamma[2-l-2i\varpi]}\sim 2i(-1)^{l+1}\frac{l+1}{l(l-1)}\varpi
\prod_{k=1}^l(k^2+4\varpi ^2)\,, \end{eqnarray}
to simplify the LHS of (\ref{eigenfin}) to
\begin{eqnarray} LHS^{-1}=\left (\frac{(l-1)!}{(2l)!}\right )^2\frac{2i(l+1)(r_+-r_-)^{-2l-1}}{(l-1)(2l+1)}\varpi
\prod_{k=1}^l(k^2+4\varpi ^2)\equiv iP_1\varpi\,. \label{LHS}
\end{eqnarray}
Here $P_1$ is obviously positive. The RHS of (\ref{eigenfin}) can
be simplified by noting that
$ \Gamma[-n+\epsilon] \sim \frac{(-1)^n}{n!\epsilon}\,,\quad \epsilon \rightarrow 0$.
We then have
\beq
RHS&=&-\frac{(l+1)(l+2)}{l(l-1)}\frac{\Gamma[1/2-l]\Gamma[2+l-\omega]\Gamma[2+l+\omega]}
{\Gamma[3/2+l]\Gamma[3/2-\omega]\Gamma[3/2+\omega]}=\\& &
-\frac{(l+1)(l+2)}{l(l-1)}\frac{\Gamma[1/2-l]\Gamma[-2n-i\delta]
\Gamma[2n+2l+4+i\delta]}
{\Gamma[3/2+l]\Gamma[-2n-l-1/2-i\delta]\Gamma[2n+l+7/2+i\delta]}\cong
iP_2\frac{1}{\delta}\,, \label{RHS} \eeq
with $P_2$ a positive constant,
\begin{eqnarray} P_2\equiv -\frac{(l+1)(l+2)}{l(l-1)}\frac{\Gamma[2n+2l+4]\Gamma[1/2-l]}
{(2n)!\Gamma[3/2+l]\Gamma[2n+l+7/2]\Gamma[-2n-l-1/2]}
\,.\end{eqnarray}
Finally, equating the RHS and LHS of (\ref{eigenfin}), i.e., the
inverse of  (\ref{LHS}) with (\ref{RHS}) we get
\begin{eqnarray} \delta=-P_1P_2\varpi\,. \end{eqnarray}
Clearly, there is an instability whenever $\varpi$ is negative
(superradiant regime), since $\Psi\propto e^{i\omega t}\propto
e^{-\delta t} $.

\section{\label{sec:endpoint} The endpoint state of the classical instability}
It is in general impossible to predict the endpoint of an unstable
system but, at least in the Kerr-AdS case, one can actually evolve
the system perturbatively, or at the least make some educated
guesses. The black hole is evolving while the instability is
acting: the angular momentum of the hole is decreasing since
supperadiant amplification of modes outside of the horizon will
transfer angular momentum to the exterior fields. The energy
release through gravitational waves (or any other field $\Psi \sim
\Psi_0e^{{\rm Im}[\omega]t}\sin{{\rm Re}[\omega]t}$), goes like
$dE/dt \sim |\dot{\Psi}|^2$ and therefore between $t=0$ and $t=T$
we have a radiated energy of
\begin{eqnarray} E \sim |\Psi_0|^2 e^{2T {\rm Im}[\omega]}
\frac{{\rm Re}[\omega]^2}{2{\rm Im}[\omega]}\sim
|\Psi(T)|^2\frac{{\rm Re}[\omega]^2}{2{\rm Im}[\omega]}
\,.\end{eqnarray}
Now, an upper limit estimate for the total energy radiated can be
obtained by equating to zero the final angular momentum and
assuming an adiabatic process. Under these conditions the first
law yields, for $a,r_+\ll \ell$,
\begin{eqnarray} \Delta M \sim \frac{a^2}{2 r_+}\,. \end{eqnarray}
We thus conclude that for
\begin{eqnarray} \frac{a^2}{r_+} \frac{{\rm Im}[\omega]}{{\rm Re}[\omega]^2}\ll 1\,, \label{constraint}\end{eqnarray}
the amplitude $\Psi$ of the field is never large, i.e., that we can trust the perturbative regime at all times,
since the instability stops for sufficiently low rotation. It is of course, possible that the instability
timescale is much lower than the other timescales at stake (like for instance, a rotation period) that the
instability is effectively unimportant, but we shall assume otherwise. Working in regime (\ref{constraint}) we
should then be able to follow the evolution of the instability in some detail, as we now explain.

From the results of the last section and from \cite{AdSbomb} one has that the real part of the frequencies that
can propagate in the Kerr-AdS background are
\begin{eqnarray} \omega= \frac{2n+l+c_s}{\ell}\,,
\label{real freq}
\end{eqnarray}
where $c_s$ is a constant that depends on the spin of the
particular wave we are considering; e.g., for scalar modes one has
$c_s=3$ \cite{AdSbomb}, and for gravitational modes one has
$c_s=2$ [see \eqref{adsfreq}]. The condition that must be
satisfied in order to have superradiance is $\omega\leq m
\Omega_{\rm H}$. Use of (\ref{real freq}) with large $l= m$,
together with $a\leq \ell$, yields that superradiance holds while
\begin{eqnarray}
r_+ \lesssim r_+^{\rm c} \qquad {\rm with} \qquad r_+^{\rm
c}=\sqrt{a \ell} \,.
 \label{superrad holds}
\end{eqnarray}

The superradiant amplification of the wave is fed by the rotational energy of the black hole and thus, as the
superradiant scattering proceeds, the angular parameter $a$ decreases. Classically, when $a$ decreases the
radius of the horizon, $r_+$, increases. Indeed, the variation of the horizon area (\ref{area}) yields
\begin{eqnarray}
dA=\frac{8\pi}{\Sigma}\left[ r_+ dr_+ + \frac{r_+^2+\ell^2}{\ell^2 \Sigma}a da \right]\,,
  \label{variation area}
\end{eqnarray}
and the classical constraint $dA\geq 0$ implies that $dr_+ >0$ when $da<0$. This fact, together with condition
(\ref{superrad holds}), allows us to clearly identify the endpoint of the superradiant classical instability.
We start with a small black hole with small $r_+$ and an angular parameter $a$ satisfying the inequality
displayed in (\ref{superrad holds}). As the superradiant scattering continues, $a$ decreases and $r_+$
increases. But as $a$ decreases the critical radius $r_+^{\rm c}=\sqrt{a \ell}$ also decreases, and thus there
is a critical minimum value of $a$ for which the radius of the black hole reaches the value of the critical
radius $r_+^{\rm c}$. This critical point describes the endpoint of the superradiant classical instability. We
end up with a Kerr-AdS black hole whose horizon radius is larger than the initial black hole and with smaller
rotation. This black hole is in equilibrium with rotating radiation that rotates in the same sense as the black
hole. This radiation is trapped between the black hole horizon and the effective AdS wall.

But we can identify another important property of this final
configuration. First note that the coordinate transformation
\begin{eqnarray}
& & \varphi=\phi-\frac{a}{\ell^2}t \nonumber \\
& & \varrho \cos \Theta=r\cos\theta \nonumber \\
& &  \varrho^2= \frac{1}{\Sigma}\left(
r^2\Delta_{\theta}+a^2\sin^2\theta \right)\,,
 \label{coord transf}
\end{eqnarray}
takes the Kerr-AdS metric (\ref{metric}) into a form in which the
asymptotic AdS nature of the Kerr-AdS solution is manifest. In
particular, in these coordinates the angular velocity of the
horizon is
\begin{eqnarray}
\Omega=\frac{a (1+ r_+^2/\ell^2)}{r_+^2+a^2} \,,
  \label{Omega termod}
\end{eqnarray}
and the angular velocity of the spacetime as $\varrho \rightarrow
\infty$ vanishes. The relation between the angular velocities at
the horizon in the two coordinate systems,
$(t,\varrho,\Theta,\varphi)$ and $(t,r,\theta,\phi)$, is
 $\Omega=\Omega_{\rm H}+a/\ell^2$.
In these new coordinates the $M=0$ Kerr AdS metric, that describes
the asymptotic behavior of the Kerr AdS metric, takes the standard
form,
\begin{eqnarray}
ds^2=-\left ( 1+\frac{\varrho^2}{\ell^2}\right )dt^2 +\left (
1+\frac{\varrho^2}{\ell^2}\right )^{-1}\,d\varrho ^2
+\varrho^2\,\left (d\Theta^2 + \sin^2\Theta \,d\varphi \right )\,.
 \label{AdS metric}
\end{eqnarray}
A hypersurface of constant large radius $\varrho$ in the Kerr AdS
has therefore a metric that is conformal to a three dimensional
Einstein universe,
\begin{eqnarray}
ds^2_{\rm boundary}=\varrho^2 \left ( -\ell^{-2}dt^2 +d\Theta^2 +
\sin^2\Theta \,d\varphi \right )\,,
 \label{conf einstein univ}
\end{eqnarray}
The coordinates $t$ and $\varphi$ in (\ref{conf einstein univ})
are constrained to an identification required by regularity.
Indeed, when we consider the Euclidean section of the Kerr-AdS
black hole by analytically continuing the time coordinate and the
angular parameter $a$, one finds that to avoid a conical
singularity at $\varrho_+$ one must identify the Euclidean time
coordinate with period $\beta=1/T$ where $T$ is the black hole
temperature (\ref{temperature}). But this identification must also
be accompanied by a rotation in $\varphi$, i.e., we must identify
the points $(t,\varrho,\Theta,\varphi)\sim
(t+i\beta,\varrho,\Theta,\varphi+i\beta\Omega)$, with $\Omega$
given by (\ref{Omega termod}). Now, the boundary solution
(\ref{conf einstein univ}) inherits the above identifications from
the bulk geometry. In particular this means that this boundary
Einstein universe is rotating with an angular velocity $\Omega$
given by (\ref{Omega termod}). Moreover, this solution is the
boundary of a bulk AdS background whose typical radius is
 $R_{\rm AdS}=\ell$, and thus the linear velocity of the rotating
 Einstein universe is $v=\Omega \ell$. Therefore, for $\Omega>
 1/\ell$ ($\Omega<1/\ell$), the Einstein universe on the boundary rotates faster (slower) than
 the speed of light. At the critical angular velocity
 $\Omega=1/\ell$ the boundary rotates exactly with light velocity.
 Use of (\ref{Omega termod}) allows us to write the equivalent
 condition
\begin{eqnarray}
\Omega= \frac{1}{\ell}  \quad \Leftrightarrow  \quad r_+= \sqrt{a
\ell} \,.
  \label{light veloc cond}
\end{eqnarray}
But this value of $r_+$ coincides with the critical value $r_+^{\rm c}$ defined in (\ref{superrad holds}).

To conclude, we find that the endpoint of the classical instability is described by a Kerr-AdS black hole whose
boundary is an Einstein universe rotating with the light velocity. This black hole should be slightly oblate
since the most unstable modes are those with $l=m=2$ and it should co-exist in equilibrium with a certain
amount of outside radiation.

\section{\label{conc}Conclusion}
We have shown that small Kerr-AdS black holes rotating with a velocity higher than the velocity of light are
unstable: any small perturbation is exponentially amplified, via a superradiant mechanism, and the system
behaves as a black hole bomb \cite{blackbomb}. We have computed the growing timescales and oscillation
frequencies of the corresponding unstable modes. More importantly, we have shown that for a large class of
these geometries one can follow the evolution of the system, as the perturbative regime is always valid. The
endpoint of the instability corresponds to a Kerr-AdS black hole whose boundary is an Einstein universe
rotating with the light velocity. This black hole is expected to be slightly oblate and to co-exist in
equilibrium with a certain amount of outside radiation. It is thus conceivable that new solutions to Einstein's
equations exist which describe such a geometry. In fact, the perturbative approach followed here already gives
such a geometry to first order in the ``oblatness'' parameter, i.e., in deviation from the Kerr-AdS black hole
solution.

\section*{Acknowledgements}
We thank Ted Jacobson for useful discussions regarding the issue of the final state for this instability.
Research at the Perimeter Institute is supported in part by funds from NSERC of Canada and MEDT of Ontario.
O.D. acknowledges financial support from Funda\c c\~ao para a Ci\^encia e Tecnologia (FCT) - Portugal through
grant SFRH/BPD/2004, and the support of a NSERC Discovery grant through the University of Waterloo. S.Y. is
supported by the Grant-in-Aid for the 21st Century COE ``Holistic Research and Education Center for Physics of
Self-organization Systems'' from the ministry of Education, Science, Sports, Technology, and Culture of Japan.

\appendix
\section{\label{app:instddim} The instability in an arbitrary number of dimensions}
In this appendix we will briefly sketch how to extend the results
in the body of the paper to higher dimensions. In particular, we
consider the subset of gravitational perturbations considered in
\cite{KunduriLuciettiReall}. There it is shown that a special
subset of gravitational perturbations obeys the equation:

\be -\frac{rf}{h}\frac{d}{dr}\left (\frac{rf}{h}\frac{d\Psi}{dr}\right )+V\Psi=0\,, \ee
where
\begin{eqnarray} V=\frac{r^2f\sqrt{h}}{h^2r^{N+1}}\frac{d}{dr}\left
[\frac{rf}{h}\frac{d}{dr}(\sqrt{h}r^N)\right]+\frac{r^2f}{h^2}\mu^2-(\omega-m\Omega)^2
+\frac{f}{h^2}\left[ l(l+2N)-m^2\left (1-\frac{r^2}{h^2}\right )
+4(1-\sigma)\left (\frac{h^2}{r^2}-1\right )\right ]
\,,\end{eqnarray}
and
\begin{eqnarray} h=r\sqrt{1+\frac{2Ma^2}{r^{2N+2}}}\,,\quad f=1+r^2-\frac{2M(1-a^2)}{r^{2N}}+\frac{2Ma^2}{r^{2N+2}}\,.\end{eqnarray}
The quantity $\sigma=\pm 1$, but it will have no bearing on the final result, since it drops out in the regime
we focus on. We will set the cosmological radius $\ell=1$ and so that all quantities are dimensionless. We
again consider the regime $r_+\ll1$ and $r_+^2\ll a \ll r_+$. In this regime we have
\beq 2M &\sim& r_+^{2N}\,,\\
h&\sim&r \,,\\
\Omega&\sim & \frac{a}{r_+^2}\gg 1\,. \eeq
The wave equation thus reduces to
\begin{eqnarray} f\frac{d}{dr}\left (f\frac{d\Psi}{dr}\right )+\left [(\omega-m\Omega)^2-f\left ( \frac{l(l+2N )}{r^2} +
\frac{(2N+1)}{4r^2}\left [(2N-1)f+2rf'\right ]+\mu^2 \right
)\right ]\Psi=0\,, \end{eqnarray}

If we change wavefunction by defining $\Psi=r^{\beta^0/2}Z(r)$, with $\beta^0=2N+1$, we get
\begin{eqnarray} \frac{f}{r^{\beta^0}}\partial_r\left (fr^{\beta^0}\partial_r Z\right )+\left [ \left (\omega-m\Omega\right
)^2-f\left (\frac{l(l+2N)}{r^2}+\mu^2\right )\right]Z=0
\,.\label{nr}\end{eqnarray}
%
\subsection{Solution in the near-region}
In the near region, $f\sim 1-\frac{2M}{r^{2N}}$ so it is natural to introduce the following change of variables
\begin{eqnarray} v=1-\frac{2M}{r^{2N}}\,, \end{eqnarray}
Then, Eq.\ (\ref{nr}) reads
\begin{eqnarray} (1-v)^2v^2\partial^2_vZ+v(1-v)^2\partial_v Z+\left ((\omega
-m\Omega_H)^2\frac{r_+^2}{4N^2}-\frac{v}{4N^2}\left
[l(l+2N)+\mu^2r_+^2\right ] \right )Z=0\,, \end{eqnarray}

This equation can be put in a standard hypergeometric form by setting
\beq
Z&=&v^{c_1}(1-v)^{c_2}{\cal F}\,,\\ c_1&=&-ia\,,\\
c_2&=&\frac{1}{2}\left (1+\sqrt{1-4(a^2-b)}\right )\sim  \frac{1}{2}\left (1+\frac{N+l}{N}\right )\,,\\
a&=&(\omega-m\Omega)\frac{r_+}{2N}\,,\\
b&=&\frac{1}{4N^2}\left [ l(l+2N)+\mu^2r_+^2\right ]\,.\eeq
The result is
\begin{eqnarray} v(1-v)\partial^2_v {\cal F}+\left [\gamma-v(1+\alpha+\beta) \right ]\partial_v{\cal F}-\alpha\beta {\cal
F}=0\,, \label{hypergeom3} \end{eqnarray}
where
\beq \gamma=1-2ia\,,\quad \alpha=\beta=c_1+c_2\,. \eeq
The most general solution of Eq.\ (\ref{hypergeom3}) in the neighborhood of $v=0$ is
\begin{eqnarray} {\cal F}=A\,F(\alpha,\beta,\gamma,v)+ B\,v^{1-\gamma} F(\alpha-\gamma+1,\beta-\gamma+1,2-\gamma,v)\,. \end{eqnarray}
Since the second term describes an outgoing wave near the horizon, we set $B=0$. The asymptotic behavior of the
near-horizon solution is
\begin{eqnarray} Z \sim r^{l}(2M)^{-l/(2N)}
\frac{\Gamma(\gamma)\Gamma(\alpha+\beta-\gamma)}{\Gamma(\alpha)\Gamma(\beta)}
+r^{-2N-l}(2M)^{1+l/(2N)}\frac{\Gamma(\gamma)\Gamma(\gamma-\alpha-\beta)}{\Gamma(\gamma-\alpha)
\Gamma(\gamma-\beta)}\,, \label{match1} \end{eqnarray}
where we have used the property of the hypergeometric functions:
\begin{eqnarray}
& \hspace{-2cm} F(\alpha,\beta,\gamma,v)=
(1\!-\!v)^{\gamma-\alpha-\beta}
\frac{\Gamma(\gamma)\Gamma(\alpha+\beta-\gamma)}{\Gamma(\alpha)\Gamma(\beta)}
\,F(\gamma-\alpha,\gamma-\beta,\gamma\!-\!\alpha\!-\!\beta\!+\!1,1\!-\!v)
& \nonumber \\ & \hspace{-0.2cm}+
\frac{\Gamma(\gamma)\Gamma(\gamma-\alpha-\beta)}{\Gamma(\gamma-\alpha)\Gamma(\gamma-\beta)}
\,F(\alpha,\beta,-\gamma\!+\!\alpha\!+\!\beta\!+\!1,1\!-\!v)\,.
\label{transformation law2}
\end{eqnarray}
%

\subsection{Solution in the far-region}
In the far-region, $r\gg M$, the effects induced by the black hole can be neglected ($a\sim 0$, $M \sim 0$,
$\Delta_r \sim r^2[1+r^2/\ell^2]$) and the radial wave equation (\ref{nr}) reduces to the wave equation of a
scalar field of frequency $\omega$ and angular momentum $l$ in a pure AdS background. The wave equation can be
written in a standard hypergeometric form. First we introduce a new radial coordinate,
\begin{eqnarray}
x^{-1}=1+\frac{r^2}{\ell^2}\, , \qquad 0\leq x \leq 1\,,
 \label{new far radial coordinate}
\end{eqnarray}
with the origin of the AdS space, $r=0$, being at $x=1$, and $r=\infty$ corresponds to $x=0$. Then the radial
wave equation (\ref{nr}) can be written as
\begin{eqnarray}
& & \hspace{-0.5cm} x(x\!-\!1)\partial_x^2 Z+(x+N) \,\partial_x Z- \!\left [ \frac{\omega^2}{4}
 + \frac{l(l+2N)}{4(x\!-\!1)}\right ]\! Z=0\,.
\nonumber \\
& &
 \label{far wave eq with x}
\end{eqnarray}
Through the definition
\begin{eqnarray}
Z=x^{N+1} (1-x)^{l/2}\,\Upsilon \,,
 \label{far hypergeometric function}
\end{eqnarray}
the radial wave equation becomes
\begin{eqnarray}
& &  \hspace{-0.5cm} x(1\!-\!x)\partial_x^2 \Upsilon+ {\biggl [} (2+N)-\left ( 3+l+2N+\right )\,x {\biggr ]}
\partial_x \Upsilon \nonumber \\
& & \hspace{1.5cm}- \frac{1}{4}(2+l+2N-\omega)(2+l+2N+\omega) \Upsilon=0\,.
 \label{far wave hypergeometric}
\end{eqnarray}
This wave equation is a standard hypergeometric equation \cite{abramowitz}, $x(1\!-\!x)\partial_x^2
F+[\gamma-(\alpha+\beta+1)x]\partial_x F-\alpha\beta F=0$, with
\begin{eqnarray}
& & \hspace{-0.5cm} \alpha=\frac{2+l+2N-\omega}{2} \,, \qquad \beta=\frac{2+l+2N+\omega}{2} \,, \qquad
\gamma=2+N \,. \nonumber \\
& &
 \label{far hypergeometric parameters}
\end{eqnarray}
The most general solution in the neighborhood of $x=0$ is
\be \Upsilon=A\,F(\alpha,\beta,\gamma,x)+ B\,x^{1-\gamma}
F(\alpha-\gamma+1,\beta-\gamma+1,2-\gamma,x)\,.\label{sol45} \ee
Since $F(a,b,c,0)=1$, as $x\rightarrow \infty$ this solution behaves as $Z\sim A r^{-2N-2}+B$. But the AdS
infinity behaves effectively as a wall, and thus the scalar field must vanish there which implies that we must
set $B=0$ in (\ref{sol45}).

We are now interested in the small $r$, $x\rightarrow 1$, behavior of (\ref{sol45}). We note that $\gamma$ is
an integer and that $\gamma=\alpha+\beta-(l+N)$ and thus some care must be exercised in expanding this
function. We get \cite{abramowitz}
\begin{eqnarray}
& & \hspace{-1.5cm} Z \sim C\,{\biggl [} \frac{\Gamma(1+N)\Gamma(N+2)}{\Gamma(\frac{2+l+2N-\omega}{2})
\Gamma(\frac{2+l+2N+\omega}{2})}\:
r^{-l-2N}\nonumber \\
& &
 +\frac{(-1)^{l+N}\Gamma(N+2)}
 {\Gamma(l+N+1)
\Gamma(\frac{2+l+2N-\omega}{2}-l-N)\Gamma(\frac{2+l+2N+\omega}{2}-l-N)}\: r^{l} {\biggr ]}.
 \label{far field-small r}
\end{eqnarray}
This can be matched to (\ref{match1}), and the instability timescale can be obtained using the procedure
explained in the main body of the paper.



\begin{thebibliography}{99}

\bibitem{maldacena} J. Maldacena, Adv. Theor. Math. Phys.
{\bf 2}, 231 (1998).

\bibitem{witten0} E. Witten, Adv. Theor. Math. Phys. {\bf 2}, 253
(1998).

\bibitem{maldacenareview} O. Aharony, S. S. Gubser, J. Maldacena,
H. Ooguri, Y. Oz, Physics Reports {\bf 323}, 183 (2000).

\bibitem{hawkingpage} S. W. Hawking and D. N. Page,
Commun. Math. Phys. {\bf 87}, 577 (1983).

\bibitem{aharonyetall} O. Aharony, J. Marsano, S. Minwalla, K. Papadodimas,
M. Van Raamsdonk, Phys. Rev. D{\bf 71}, 125018 (2005).

\bibitem{louwin} J. Louko and S. N. Winters-Hilt,  Phys. Rev.
D{\bf 54} (1996) 2647.

\bibitem{pele} C. S. Pe\c ca and P. S. Jos\'e Lemos,  Phys. Rev. D{\bf 59} (1999) 124007.

\bibitem{mi1} P. Mitra,  Phys. Lett. B{\bf 441} (1998) 89.

\bibitem{mi2} P. Mitra,  Phys. Lett. B{\bf 459} (1999) 119.

\bibitem{cejm1} A. Chamblin, R. Emparan, C. V. Johnson, and R. C. Myers,
Phys. Rev. D{\bf 60} (1999) 064018.

\bibitem{cejm2} A. Chamblin, R. Emparan, C. V. Johnson, and R. C. Myers,
 Phys. Rev. D{\bf 60} (1999) 104026.

\bibitem{cck} M. M. Caldarelli, G. Cognola, D. Klemm,
Class. Quantum Grav. {\bf 17} (2000) 399.

\bibitem{hr}  S. W. Hawking, and H. S. Reall, Phys. Rev. D{\bf 61} 024014
(2000).

\bibitem{berpar} D. S. Berman, M. K. Parikh, Phys. Lett. B{\bf 463} (1999) 168

\bibitem{hhtr} S. W. Hawking, C. J. Hunter, and M. M. Taylor-Robinson, Phys.
Rev. D{\bf 59}, 064005 (1999).


\bibitem{birm} D. Birmingham,  Class. Quant. Grav. {\bf 16} (1999) 1197.

\bibitem{caldklem} M. M. Caldarelli and D. Klemm,  Nucl. Phys. B{\bf 555} (1999) 157.

\bibitem{emp} R. Emparan, JHEP {\bf 06} (1999) 036.

\bibitem{teubhb} W. H. Press and S. A. Teukolsky,
Nature {\bf 238}, 211 (1972).

\bibitem{blackbomb} V. Cardoso, O. J. C. Dias, J. P. S. Lemos and S. Yoshida,
Phys. Rev. D {\bf 70}, 044039 (2004); Erratum-ibid.D {\bf 70}, 049903 (2004).

\bibitem{AdSbomb} V. Cardoso and O. J. C. Dias, Phys. Rev. D {\bf 70}, 084011 (2004).

\bibitem{hawkingreall} S. W. Hawking and H. S. Reall,
Phys. Rev. D {\bf 61}, 024014 (1999).

\bibitem{bs} V.~Cardoso and J.~P.~S.~Lemos,
Phys.\ Lett.\ B {\bf 621}, 219 (2005); V.~Cardoso and S.~Yoshida, JHEP {\bf 0507}, 009 (2005); O.~J.~C.~Dias,
Phys.\ Rev.\ D {\bf 73}, 124035 (2006).

\bibitem{KunduriLuciettiReall} H. K. Kunduri, J. Lucietti, H. S. Reall, hep-th/0606076.

\bibitem{teukolsky} S. A. Teukolsky, Astrophys. J. {\bf 185}, 635 (1973).

\bibitem{mosschamgia} C. M. Chambers and I. G. Moss, Class. Quant. Grav. {\bf 11}, 1035 (1994);
M. Giammatteo and I. G. Moss, Class. Quant. Grav. {\bf 22}, 1803 (2005).

\bibitem{cardosogpert} V. Cardoso and J. P. S. Lemos, Phys. Rev. D {\bf 64}, 084017 (2001);
V. Cardoso, R. Konoplya and J. P. S. Lemos, Phys. Rev. D {\bf 68}, 044024 (2003).

\bibitem{swsh} E. Berti, V. Cardoso and M. Casals, Phys. Rev. D {\bf 73}, 024013 (2006);
Erratum-ibid. D {\bf 73}, 109902 (2006).

\bibitem{abramowitz} M. Abramowitz and
A. Stegun, {\it Handbook of mathematical functions}, (Dover Publications, New York, 1970).

\bibitem{moss}
J. S. Avis, C. J. Isham and D. Storey, Phys. Rev. D {\bf 18}, 3565 (1978); P. Breitenlohner and D. Z. Freedman,
Phys. Lett. B {\bf 153}, 137 (1982); I. G. Moss and J. P. Norman, Class. Quant. Grav. {\bf 19}, 2323 (2002).

%
%



\end{thebibliography}
\end{document}